\documentclass[10pt,twocolumn,letterpaper]{article}
\usepackage{marvosym}
\usepackage{booktabs}
\usepackage{colortbl}

\usepackage{cvpr}              

\usepackage[dvipsnames]{xcolor}

\definecolor{cvprblue}{rgb}{0.21,0.49,0.74}
\usepackage[pagebackref,breaklinks,colorlinks,citecolor=cvprblue]{hyperref}

\def\paperID{*****} 
\def\confName{CVPR}
\def\confYear{2024}

\title{Advancing COVID-19 Detection in 3D CT Scans}

\author{Qingqiu Li$^1$, Runtian Yuan$^1$, Junlin Hou$^2$, Jilan Xu$^1$, \\
Yuejie Zhang$^1$, Rui Feng$^1$, Hao Chen$^2$\\
\\
$^1$ Fudan University\\
$^2$ The Hong Kong University of Science and Technology\\
}

\begin{document}
\maketitle
\begin{abstract}
To make a more accurate diagnosis of COVID-19, we propose a straightforward yet effective model. Firstly, we analyse the characteristics of 3D CT scans and remove the non-lung parts, facilitating the model to focus on lesion-related areas and reducing computational cost. We use ResNeSt50 as the strong feature extractor, initializing it with pretrained weights which have COVID-19-specific prior knowledge. Our model achieves a Macro F1 Score of 0.94 on the validation set of the 4th COV19D Competition Challenge \uppercase\expandafter{\romannumeral1}, surpassing the baseline by 16\%.  This indicates its effectiveness in distinguishing between COVID-19 and non-COVID-19 cases, making it a robust method for COVID-19 detection.
\end{abstract}    
\section{Introduction}
\label{sec:intro}

\begin{figure}[t]
    \centering
    \includegraphics[width=1\linewidth]{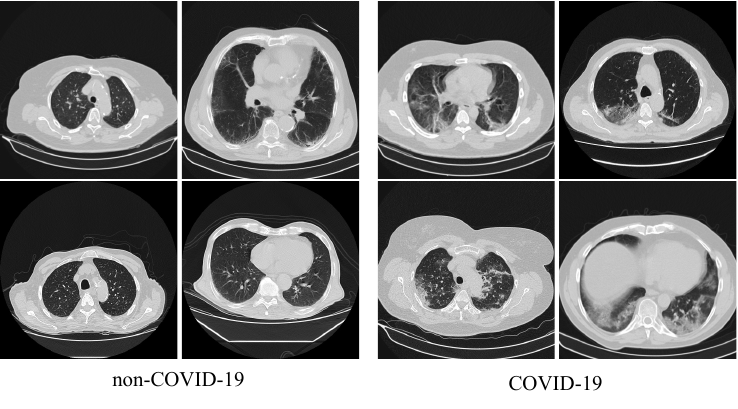}
    \caption{Samples of non-COVID-19 and COVID-19 from the COV19-CT-DB database.}
    \label{fig:arch}
\end{figure}

The outbreak of COVID-19 has led to widespread health crises and fatalities. Early detection is crucial for controlling and preventing the spread of the virus. As shown in Fig.~\ref{fig:arch}, Chest CT scans have been extensively utilized for diagnosing and monitoring COVID-19 patients, due to their ability to provide detailed insights into lung involvement's extent and severity. However, the vast number of CT images generated necessitates a significant workload for radiologists and medical practitioners, making the diagnosis process challenging.

In recent years, deep learning has been widely applied in the automatic detection of COVID-19~\cite{kollias2020deep,kollias2021mia,kollias2023deep,hou2021periphery,subramanian2022review,hou2022cmc_v2,hou2022boosting}. However, previous approaches often decompose 3D CT scans into individual 2D slices for analysis or rely on simple architectures to train models from scratch, leading to less-than-ideal classification results. To address the mentioned issues, we treat each complete CT scan as a 3D volume, removing non-lung parts that do not contribute to COVID-19 detection. Then, we utilize ResNeSt50 \cite{resnest} as the backbone and initialize it with pretrained weights provided by CMC v1 \cite{hou2021cmc}. Finally, we employ cross-entropy loss for binary classification training.

Our primary contributions are outlined as follows:
\begin{enumerate}
  \item We analyse the characteristics of 3D CT scans and remove the non-lung parts, focusing the model on lesion-related areas and reducing computational cost.
  \item We use ResNeSt50 as a strong feature extractor, initializing it with pretrained weights that have COVID-19-specific prior knowledge.
  \item Our model achieves a Macro F1 score of 0.94 on the validation set of the 4th COV19D Competition Challenge \uppercase\expandafter{\romannumeral1}, surpassing the baseline by 16\%.
\end{enumerate}

\section{Methodology}
\label{sec:method}

\begin{figure*}
    \centering
    \includegraphics[width=1\linewidth]{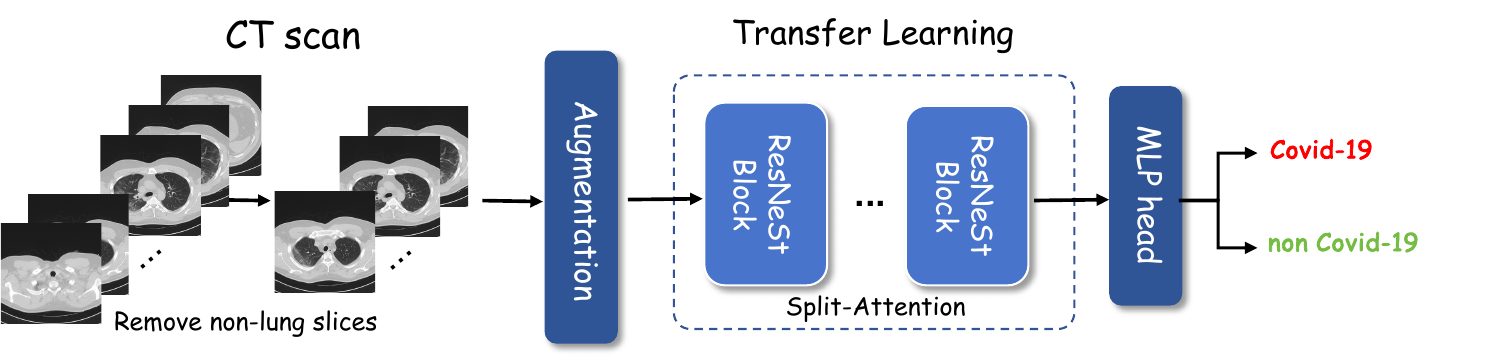}
    \caption{Overview of our framework for COVID-19 detection.}
    \label{fig:arch1}
\end{figure*}

The overall framework of our model is shown in Fig.~\ref{fig:arch1}. Firstly, we analyse the 3D CT scans and  identify the slices that are irrelevant for COVID-19 detection, i.e., the neck area at the start and the abdominal region towards the end. Consequently, we remove these slices from the volume that lacked lung regions, directing the model's focus towards areas relevant to lesion detection and simultaneously reducing computational requirements.

After processing the 3D CT scans, we chose ResNeSt50 as our feature extractor, which presents a modular split-attention block within the individual network blocks to enable attention across feature-map groups. Considering that training a model from scratch leads to poor results, we employed transfer learning. However, generic pretrained weights based on ImageNet fail to meet the specific needs of COVID-19 classification, which demands specialized medical domain knowledge. Therefore, we initialized our model with weights from CMC v1 to incorporate COVID-19-specific prior knowledge. Finally, the cross-entropy loss is employed for binary classification training.

\section{Datasets}
We evaluate our proposed approach on the COV19-CT-Database (COV19-CT-DB) ~\cite{kollias2023deep}. The COV19-CT-DB contains chest CT scans, collected in various medical centers. The database includes 7,756 3D CT scans, where 1,661 are COVlD-19 samples, whilst 6,095 refer to non COVlD-19 ones. In total, 724,273 slices correspond to the CT scans of the COVID-19 category and 1,775,727 slices correspond to the non-COVID-19 category class~\cite{arsenos2022large,arsenos2023data,kollias2020deep,kollias2020transparent,kollias2021mia,kollias2022ai,kollias2023ai}.

For Challenge \uppercase\expandafter{\romannumeral1}, the training set contains 1358 3D CT scans (655 non-COVID-19 cases and 703 COVID-19 cases). Based on this, we further enrich our training set with annotated data from Challenge \uppercase\expandafter{\romannumeral2} (120 non-COVID-19 cases and 120 COVID-19 cases), aiming to enhance the model's learning capacity and its ability to generalize across diverse cases of COVID-19 detection. The validation set consists of 326 3D CT scans (170 non-COVID-19 cases and 156 COVID-19 cases). The testing set includes 1413 scans and the labels are not available during the challenge.
\section{Experiments}
\subsection{Data Pre-Processing}
Our data pre-processing procedure is as follows. All 2D chest CT scan series are composed into a 3D volume of shape $(D,H,W)$, where $D,H,W$ denotes the number of slice, height, and width, respectively. Then, each 3D volume is resized to dimensions of (128, 256, 256). Finally, we transform the CT volume to the interval [0, 1] for intensity normalization.

\subsection{Implementation Details}
We utilize 3D ResNeSt50 as the backbone of our model. For training, data augmentations include random resized cropping on the transverse plane, random cropping on the vertical section to 64, rotation, and color jittering. We use Adam algorithm\cite{adam} as our optimizer, setting the learning rate to $1e-4$ and the weight decay to $1e-5$. Our model is trained 100 epochs on 4 RTX 3090 GPUs with a batch size of 2 per GPU. Macro F1 score is the evaluation metric, which calculates the F1 score for each category separately and then averages these scores to assess overall performance.

\subsection{Experimental Results}
Table~\ref{tab:exp} compares the classification outcomes on the validation set of Challenge \uppercase\expandafter{\romannumeral1} between the baseline model and ours. Our model's Macro F1 score surpasses the baseline by 16\% , underscoring its efficacy and establishing it as a robust baseline for COVID-19 detection.
\begin{table}[t]
\centering
\caption{Comparison results on the validation set of Challenge \uppercase\expandafter{\romannumeral1}.}
\begin{tabular}{cc}
\specialrule{\heavyrulewidth}{0pt}{0pt}
Method & Macro F1 Score \\
\hline
\hline
Baseline \cite{kollias2024domain} & 0.78 \\
\rowcolor{gray!20} \textbf{Ours} & \textbf{0.94} \\
\specialrule{\heavyrulewidth}{0pt}{0pt}
\end{tabular}
\label{tab:exp}
\end{table}
\section{Conclusion}
In this paper, we propose a straightforward yet effective model for COVID-19 detection. Firstly, we analyse the characteristics of 3D CT scans, removing non-lung regions from the entire volume. This approach not only facilitates the model to focus on lesion-related areas but also reduces computational cost.
We choose ResNeSt50 as the feature extractor, utilizing transfer learning instead of training the model from scratch. We initialized our model with pretrained weights from CMC v1 to incorporate COVID-19-specific prior knowledge. Finally, we employed cross-entropy loss for binary classification training. Based on the aforementioned techniques, our model achieves an Macro F1 Score of 0.94 on the validation set of the 4th COV19D Competition Challenge \uppercase\expandafter{\romannumeral1}, surpassing the baseline by 16\%.  
{
    \small
    \bibliographystyle{ieeenat_fullname}
    \bibliography{main}
}


\end{document}